\documentstyle[buckow]{article}

\def\beq{\begin{equation}}
\def\eeq{\end{equation}}

\begin {document}

\def\titleline{On gravitational interaction of fermions
}
\def\authors{Yuri N. Obukhov}
\def\addresses{Department of Theoretical Physics \\
Moscow State University\\
117234 Moscow, Russia}

\def\abstracttext{
We discuss some aspects of the gravitational interaction of the 
relativistic quantum particles with spin 1/2. The exact Foldy-Wouthuysen 
transformation is constructed for the Dirac particle coupled to the static 
spacetime metric. The quasi-relativistic limit of the theory is then 
analyzed. Using the analogous method, we obtain the exact Cini-Touschek
transformation and discuss the ultra-relativistic limit of the fermion
theory. We show that the Foldy-Wouthuysen transformation is not uniquely
defined, and the corresponding ambiguity is deeply rooted in the 
relativistic quantum theory.}
\large
\makefront

\section{Introduction}

The recent development of experimental technique, in particular of the 
neutron interferometric methods \cite{cow}, has provided the first direct 
tests of the interaction of the quantum spinless particles with the classical
gravitational field. There is a little doubt that a further technological 
progress (using the polarized neutrons, atomic interferometers, etc) will 
soon make it possible to measure the higher order gravitational and inertial
effects of the quantum particle {\it with} spin. Theoretical studies of
the relativistic quantum theory in a curved spacetime have predicted a
number of interesting manifestations of the spin-gravity coupling for 
the Dirac fermion, see \cite{early,fish,heni,spin}, e.g. In most cases,
the various approximate schemes were used for the case of the weak 
gravitational field. Here the exact results for an arbitrary static 
spacetime geometry are reported. 

A massive quantum particle with spin $1/2$ is described by the relativistic
Dirac theory. In the curved spacetime, the fermion wave function -- 4-spinor 
field $\psi$ -- satisfies the covariant Dirac equation
\begin{equation}
\left(i\hbar \gamma^\alpha D_\alpha - mc\right)\psi = 0\label{Dirac0}.
\end{equation}
The spinor covariant derivative is defined by
\begin{equation}
D_\alpha = h_\alpha^i D_i,\qquad 
D_i := \partial_i + {\frac i 4}\,\widehat{\sigma}_{\alpha\beta}
\,\Gamma_i{}^{\alpha\beta},\label{Dspinor}
\end{equation}
which shows that the gravitational and inertial effects are encoded in
the coframe (vierbein) and the Lorentz connection coefficients
$h_i{}^\alpha,\,\Gamma_i{}^{\alpha\beta}= -\,\Gamma_i{}^{\beta\alpha}$. 
We use the Greek alphabet for the indices which label the components
with respect to a local Lorentz frame $e_\alpha = h_\alpha^i\partial_i$,
whereas the Latin indices refer to the local spacetime coordinates $x^i$. 
For the Dirac matrices, the conventions of \cite{bjorken} are used.
In particular, we have $\beta = \gamma^{\widehat{0}}, \vec{\alpha} = \beta
\vec{\gamma}$, $\gamma_5 = -i\gamma^{\widehat{0}}\gamma^{\widehat{1}}
\gamma^{\widehat{2}}\gamma^{\widehat{3}}$, $\widehat{\sigma}^{\alpha\beta}
= i\gamma^{[\alpha}\gamma^{\beta]}$. The spin matrix is defined by 
$\vec{\Sigma} = i\vec{\gamma}\times\vec{\gamma}/2 = -\gamma_5\vec{\alpha}$.

Let us consider the metric of the static spacetime
\begin{equation}
ds^2 = V^2\,(dx^0)^2 - W^2\,(d\vec{x}\cdot d\vec{x}).\label{metVW}
\end{equation}
Here $x^0 = ct$, and $V = V(\vec{x}), W = W(\vec{x})$ are arbitrary 
functions of $\vec{x}$. Many important particular 
cases belong to this family: (i) the flat Minkowski spacetime in {\it 
accelerated frame} corresponds to the choice
$V = 1 + {(\vec{a}\cdot\vec{x})}/{c^2},\,W = 1$,
(ii) {\it Schwarzschild} spacetime in the isotropic coordinates 
with $r := \sqrt{\vec{x}\cdot\vec{x}}$ is obtained for 
$V = \left(1 - {\frac {GM}{2c^2r}}\right)\left(1 + {\frac {GM} {2c^2r}}
\right)^{-1},\, W = \left(1 + {\frac {GM}{2c^2r}}\right)^2$,
(iii) {\it de Sitter} spacetime in static frame is recovered for
$V = {\frac {1 + r^2/\ell^2}{1 - r^2/\ell^2}},\,W = 2/(1 - r^2/\ell^2)$,
where $\ell$ is the constant curvature radius (with the curvature 2-form 
$R^{\alpha\beta} = 1/\ell^2\,\vartheta^\alpha\wedge\vartheta^\beta$),
(iv) the {\it product} spacetime $R\times S^3$ (fermion on a sphere)
arises when $V = 1,\,W = \left(1 + r^2/(4L^2)\right)^{-1}$ with 
$L$ radius of the sphere $S^3$. 

Using (\ref{metVW}), one can bring the Dirac equation (\ref{Dirac0}) to
the Schr\"odinger form
\begin{equation}\label{Dirac1}
i\hbar{\frac {\partial\psi} {\partial t}} = \widehat{\cal H}\,\psi
\end{equation}
with the Hamilton operator 
\begin{equation}\label{Hamilton1}
\widehat{\cal H} = \beta mc^2V + {\frac c 2}\left[(\vec{\alpha}
\cdot\vec{p}){\cal F} + {\cal F}(\vec{\alpha}\cdot\vec{p})\right].
\end{equation}
Here we introduced ${\cal F} = V/W$.

\section{Foldy-Wouthuysen transformation}

In order to reveal the true physical content of the theory and to find
its correct interpretation, one needs to perform the Foldy-Wouthuysen (FW)
transformation \cite{FW}. Technically, this yields the representation in 
which the quantum states with positive and negative energy become uncoupled. 

We use here the approach of Eriksen \cite{erik} to construct the exact
Foldy-Wouthuysen transformation. The energy sign operator (Pauli) is defined 
by $\widehat{\Lambda} = {\widehat{\cal H}}/{\sqrt{{\widehat{\cal H}}^2}}$.
It is Hermitian, unitary, and idempotent: $\widehat{\Lambda}^2 = 
\widehat{\Lambda}^\dagger \widehat{\Lambda} = 1$. The unitary operator $U$ 
which maps the Dirac representation to the FW-representation 
\begin{equation}
\psi\longrightarrow\psi^F = U\psi,
\end{equation}
should satisfy the condition
\begin{equation}
U\,\widehat{\Lambda}\,U^\dagger = \beta.\label{UU}
\end{equation}

Remarkably, for our case, the {\it exact} FW-trans\-formation exists.
Consider the operator
\begin{equation}
J := i\gamma_5\beta = \left(\begin{array}{cc} 0&i\\ 
-i& 0\end{array}\right). 
\end{equation}
It is Hermitian, $J^\dagger = J$, unitary, and idempotent: $JJ^\dagger 
= J^2 = 1$, and it anticommutes both with the Hamiltonian, and with $\beta$:
\begin{equation}
J\widehat{\cal H} + \widehat{\cal H}J = 0,\quad 
J\beta + \beta J = 0.
\end{equation}
Then the FW transformation (\ref{UU}) is realized by 
\begin{equation}
U = {\frac 1 2}\left(1 + \beta J\right)
\left(1 + J\,\widehat{\Lambda}\right)
\end{equation}
and the corresponding FW Hamiltonian reads
\begin{equation}
\widehat{\cal H}^F = U\widehat{\cal H}U^\dagger 
= \left[\sqrt{{\widehat{\cal H}}^2}\;\right]
\!\beta + \left\{\sqrt{{\widehat{\cal H}}^2}\right\}\!J.
\end{equation}
As usual, even and odd parts of an operator $Q$ are defined by
$[Q] := {\frac 1 2}\left(Q + \beta Q\beta\right)$ and
$\{Q\} := {\frac 1 2}\left(Q - \beta Q\beta\right)$.
Explicitly, we have for the square of (\ref{Hamilton1})
\begin{equation}
\widehat{\cal H}^2 = m^2c^4\,V^2 + {\cal F}c^2p^2{\cal F} + 
{\frac {\hbar^2c^2}2}\left({\cal F}\,\vec{\nabla}\vec{f} 
- \vec{f}^2/2\right) +\,\hbar c^2\,{\cal F}\,\vec{\Sigma}\cdot
\left(\vec{f}\times\vec{p} + J\,mc\,\vec{\phi}\,\right)
\end{equation}
which contains the {\it odd} piece (the last term). Here: 
$\vec{\phi} := \vec{\nabla}V,\,\vec{f}:= \vec{\nabla}{\cal F}$.

The obtained FW Hamiltonian is {\it exact}. However, for most practical 
purposes, it is sufficient to consider the non-relativistic limit, and use 
the quasi-relativistic wave functions, treating all the interaction terms as 
perturbations. The quasi-relativistic approximation is straightforwardly 
obtained by assuming that $mc^2$ term is dominating, and expanding 
$\sqrt{{\widehat{\cal H}}^2}$ in powers of $1/(mc^2)$. 
However, the massless case is covered only by the exact result.

Expanding $\sqrt{{\widehat{\cal H}}^2}$ in powers of $1/(mc^2)$, 
we find finally the quasi-relativistic Hamiltonian:
\begin{eqnarray}
\widehat{\cal H}^F &\approx& \underbrace{\beta\,mc^2\,V} + 
\underbrace{ {\frac 1 {4m}}\,\beta\,
\left({\frac 1 W}p^2{\cal F} + {\cal F}p^2{\frac 1 W}\right)}
+\,\underbrace{{\frac {\hbar^2}{4mW}}\,\beta\,(\triangle{\cal F})} 
-\,{\frac {\hbar^2} {8mV}}\,\beta\,\vec{f}^2 \nonumber\\
&&\ {\rm COW}\qquad\quad {\rm kinetic\ rel.\ red\ shift}\qquad
{\rm grav.\ Darwin\ term}\label{nonH} \\
&& +\,\underbrace{{\frac {\hbar} {4m}}\,\beta\,\vec{\Sigma}\cdot
\left({\frac 1 W}\,\vec{f}\times\vec{p} + \vec{f}\times\vec{p}
\,{\frac 1 W}\right)}+\,{\frac {\hbar c}{2W}}
\,(\vec{\Sigma}\cdot\vec{\phi}).\nonumber\\
&& \qquad {\rm grav/inert.\ spin-orbit.\ term}\nonumber
\end{eqnarray}
The first two terms describe the familiar effects already measured 
experimentally for spinless particles (Colella-Overhauser-Werner and 
Bonse-Wroblewski, \cite{cow}). The first term in the second line 
represents the new inertial/gravitational spin-orbital momentum effects,
cf. \cite{fish,heni}. The ``gravitational Darwin'' term admits a physical 
interpretation similar to usual electromagnetic Darwin term, reflecting 
the zitterbewegung fluctuation of the fermion's position with the mean 
square $<\!\!(\delta r)^2\!\!>\sim\hbar^2/(mc)^2$. 

It is interesting to observe the emergence of the spin-gravitational 
moment coupling which is described by the last term in (\ref{nonH}).
Such interaction was predicted, in a phenomenological approach, by
Kobzarev and Okun \cite{gravmom} and was discussed by Peres \cite{peres},
see also the recent reviews \cite{mash}. The presence of this term 
demonstrates the validity of the equivalence principle for the Dirac 
fermions \cite{spin}. 

\section{Cini-Touschek transformation}

In the study of the high-energy neutrino effects in the gravitational 
field of a massive compact object (see \cite{konno}, e.g.), it is convenient 
to use a different representation which is directly related to the 
ultra-relativistic rather than to the quasi-relativistic limit. 
The new representation is in this sense complementary to the FW picture. 

The corresponding limit (when $mc^2 \ll c|p|$) is achieved with the help
of the Cini-Touschek (CT) transformation \cite{cini}. We can construct the 
exact CT-transformation for a fermion in the static metric (\ref{metVW}) 
using the scheme similar to the above FW case. To begin with, we observe
that the operator
\begin{equation}
\widehat{\cal P} = {\frac {\vec{\alpha}\cdot\vec{p}}{|p|}}
\end{equation}
is Hermitian, unitary, and idempotent, $\widehat{\cal P}^2 = 1$. It is  
proportional to the chirality operator $\widehat{\chi} = \vec{\Sigma}\cdot
\vec{p}/|p| = -\,\gamma_5\,\widehat{\cal P}$. Evidently, we have
\begin{equation}
\widehat{\cal P}\,J + J\,\widehat{\cal P} = 0.
\end{equation}
In complete analogy to (\ref{UU}), the CT-trans\-formation is determined by 
the unitary operator $U$ which satisfies
\begin{equation}
U\,\widehat{\Lambda}\,U^\dagger = \widehat{\cal P}.
\end{equation}
Thus, technically, we need to replace $\beta \leftrightarrow 
\widehat{\cal P}$ everywhere in the above derivations. The explicit 
CT-operator then reads
\begin{equation}
U = {\frac 1 2}\left(1 + \widehat{\cal P}\,J\right)
\left(1 + J\,\widehat{\Lambda}\right)
\end{equation}
and the Cini-Touschek Hamiltonian is 
\begin{equation}\label{CTH}
\widehat{\cal H}^{CT} = \left[\sqrt{{\widehat{\cal H}}^2}\;\right]^{\cal P}
\!\widehat{\cal P} + \left\{\sqrt{{\widehat{\cal H}}^2}\right\}^{\cal P}\!J.
\end{equation}
Here, the ``$\widehat{\cal P}$-odd/even'' parts of an operator are defined 
by the same token as the usual ``$\beta$-odd/even'' parts. As a simple
application, we check that for the free particle (\ref{CTH}) yields 
$\widehat{\cal H}^{CT} = E\,\widehat{\cal P}\approx c\vec{\alpha}\cdot
\vec{p}$ which is the correct ultra-relativistic Hamiltonian.

\section{Ambiguities}

The presence of the spin-gravitational moment in the quasi-relativistic
FW Hamiltonian requires some comments. 

FW transformation is defined with a certain ambiguity. 
Let us consider the unitary transformation $U' = e^{iS}$ with 
\begin{equation}
S = {\frac \beta {mc}}\left\{b(x)\,(\vec{\Sigma}\vec{p}) + 
(\vec{\Sigma}\vec{p})\,b(x)\right\},
\end{equation}
where $b(x)$ is an arbitrary function of the spatial coordinates. The 
spaces of quantum states with positive and negative energies are invariant
under the action of this operator. For the Hamiltonian of the unitary
equivalent representation, we find, in a perturbative manner: 
\begin{eqnarray}
\widehat{\cal H}' = U'\widehat{\cal H}^F{U'}^\dagger &=& 
\widehat{\cal H}^F + 2\hbar c\,b\,(\vec{\Sigma}\vec{\phi}) + {\frac 
{i\hbar} m}\,\beta\left[b\,(\vec{\Sigma}\vec{p}), \left(1/W + 2b\right)
(\vec{\Sigma}\vec{\phi})\right] \nonumber\\
&& + \,{\frac {\hbar^2}{2m}}\left(1/W + 2b\right)\beta
\left[(\vec{\Sigma}\vec{\nabla}b),(\vec{\Sigma}\vec{\phi})\right] 
+ {\cal O}(1/m^2).\label{newH}
\end{eqnarray}
Using (\ref{nonH}), we find that the choice $2b = -\,{\frac 1 {2W}}$ 
yields the approximate form of the FW Hamiltonian reported by Fischbach 
et al \cite{fish} and by Hehl and Ni \cite{heni}.

The mentioned ambiguity is deeply rooted in the relativistic quantum theory. 
The FW representation is often treated merely as a rigorous method to derive 
the quasi-relativistic limit of the Dirac equation (see \cite{rose}, e.g.), 
refining the ``non-rigorous'' derivation based on the direct elimination of 
the so-called small components of the 4-spinor wave function. And indeed,
one can straightforwardly verify that the direct derivation of the Pauli
equation suffers from the same ambiguity: Recall that after eliminating
the small components, the remaining 2-spinor wave function should be 
properly normalized \cite{AB}. The corresponding normalization operator
is not uniquely defined and this yields the transformation of the type 
(\ref{newH}) of the quasi-relativistic Hamiltonian.

Furthermore, one can easily find the relevant ambiguities of that kind in 
the full (relativistic) Dirac theory. For example, the Hamiltonian of the 
free particle $\widehat{\cal H} = c\,(\vec{\alpha}\vec{p}) + \beta\,mc^2$ is 
invariant under the unitary transformation of the wave function described by
\begin{equation}
U = \sqrt{\frac {E + mc^2} {2E}}\left(1 + {\frac {ic} {E + mc^2}}
\,(\vec{\Sigma}\vec{p})\,\widehat{\Lambda}\right).
\end{equation}

Operators of position, spin and energy (Hamiltonian) can have different
form in the unitary equivalent representations, and one should properly
determine them in order to analyze the physical effects. Certainly, the
observable quantities measured in experiment do not depend on the 
choice of representation.

\section{Discussion and conclusion}

Approximate scheme (see Bjorken-Drell \cite{bjorken}, e.g.) was developed 
for the case of electromagnetic coupling. As it is well known, the idea is 
to remove, order by order in $1/m$, odd terms from the Hamiltonian 
$\widehat{\cal H} = \widehat{\cal H}_1 = \beta mc^2 + {\cal E} + {\cal O}$. 
A unitary transformation $\psi_2 = U_{21}\,\psi_1$, with $U_{21} = e^{iS}$, 
yields (in the time-independent case) the perturbative construction of
the new Hamiltonian
\begin{equation}
\widehat{\cal H}_2  = U_{21}\,\widehat{\cal H}_1\,U_{21}^\dagger 
= \widehat{\cal H}_1 + \left[iS,\widehat{\cal H}_1\right]
+ {\frac 1 2}\,\left[iS,\left[iS,\widehat{\cal H}_1\right]\right]
+ {\frac 1 {3!}}\,\left[iS,\left[iS,\left[iS,\widehat{\cal H}_1\right]
\right]\right] + \dots .
\end{equation}
In electrodynamics, the odd ${\cal O}$ and even ${\cal E}$ parts do not 
depend on the mass $m$. Instead, they are proportional to {\it electromagnetic}
charge $e$, and that fact makes the approximate scheme working. For example,
choosing at the first step $S = -i\beta{\cal O}/2m$, we remove the original
odd part and find $\widehat{\cal H}_2 = \beta mc^2 + {\cal E}' + {\cal O}'$ 
where the new odd part
\begin{equation}
{\cal O}' = {\frac \beta {2m}}\,[{\cal O}, {\cal E}] 
- {\frac {{\cal O}^3} {3m^2}} 
\end{equation}
has a higher order in $1/m$ than the new even part. However, for the 
gravitational/inertial case, ${\cal E}$ is proportional to the {\it 
gravitational/inertial} charge $m$. As a result, the new odd term in 
$\widehat{\cal H}_2$ is of order $m^0$. The same happens at every step 
of the approximate scheme: the ``remaining'' even terms have the same 
order in $1/m$ as the ``removed'' odd terms. This makes the issue of 
the convergence of the approximate scheme problematic. In our approach, 
this deficiency is avoided by using the {\it exact} FW transformation. 

Here we have demonstrated how to obtain the exact FW and CT transformations
in the covariant Dirac theory. The detailed discussion of the corresponding 
applications to the specific quasi-relativistic and ultra-relativistic 
physical problems will be presented elsewhere. 

\bigskip
\noindent {\bf Acknowledgment}\ I would like to thank the Alexander von 
Humboldt Foundation for the invitation and support. 


\end{document}